\documentclass[11pt]{article}

\title{Reproducible Research: A Retrospective}

\usepackage[margin=1.25in]{geometry}
\usepackage{xcolor}

\usepackage[numbers,sort&compress]{natbib}
\bibpunct{(}{)}{,}{n}{}{;}  

\author{Roger D. Peng$^1$\and Stephanie C. Hicks$^1$}

\begin{document}

\maketitle

\noindent
$^1$Department of Biostatistics, Johns Hopkins Bloomberg School of Public Health\\

\begin{abstract}
Rapid advances in computing technology over the past few decades have spurred two extraordinary phenomena in science: large-scale and high-throughput data collection coupled with the creation and implementation of complex statistical algorithms for data analysis. Together, these two phenomena have brought about tremendous advances in scientific discovery but have also raised two serious concerns, one relatively new and one quite familiar. The complexity of modern data analyses raises questions about the \textit{reproducibility} of the analyses, meaning the ability of independent analysts to re-create the results claimed by the original authors using the original data and analysis techniques. While seemingly a straightforward concept, reproducibility of analyses is typically thwarted by the lack of availability of the data and computer code that were used in the analyses. A much more general concern is the \textit{replicability} of scientific findings, which concerns the frequency with which scientific claims are confirmed by completely independent investigations. While the concepts of reproduciblity and replicability are related, it is worth noting that they are focused on quite different goals and address different aspects of scientific progress. In this review, we will discuss the origins of reproducible research, characterize the current status of reproduciblity in public health research, and connect reproduciblity to current concerns about replicability of scientific findings. Finally, we describe a path forward for improving both the reproducibility and replicability of public health research in the future.
\end{abstract}

\clearpage

\section{Introduction}

Scientific progress has long depended on the ability of scientists to communicate to others the details of their investigations. The exact meaning of ``details of their investigations" has changed considerably over time and in recent years has been nearly impossible to describe precisely using traditional means of communication. Rapid advances in computing technology have led to large-scale and high-throughput data collection coupled with the creation and implementation of complex statistical algorithms for data analysis. In the past, it might have sufficed to describe the data collection and analysis using a few key words and high-level language. However, with today's computing-intensive research, the lack of details about the data analysis in particular can make it impossible to re-create any of the results presented in a paper~\citep{peng:2011}. Compounding these difficulties is the impracticality of describing these myriad details in traditional journal publications using natural language. To address this communication problem, a concept has emerged known as \textit{reproducible research}, which aims to provide far more precise descriptions of an investigator's work to others in the field. As such, reproducible research is an extension of the usual communications practices of scientists, adapted to the modern era.

The notion of reproducible research, which was popularized in the early 1990s, was ultimately designed to address an emerging and serious issue at the time~\citep{schwab2000making}. Results of published findings were increasingly dependent on complex computations done on powerful computers, often implementing sophisticated algorithms on large datasets. Given the importance of computing to the generation of these results, it was surprising that consumers of scientific results had no ability to inspect or examine the details of the computations being done. Traditional forms of scientific publication allowed for extended descriptions of study design and high-level analysis approaches, but low-level details about computer code, data processing pipelines, and algorithms were not prioritized and generally left in an appendix or, with the wider availability of the internet, online supplement. 

Jon Claerbout, a geophysicist at Stanford University, wrote down many of the original ideas concerning reproducibilty of computational research. His concern largely focused on developing a software system whereby the research produced by his lab could be passed on to others, including the original authors, the authors' colleagues, students, research sponsors, and the general public. He noted in particular the benefits of reproducibility to the original authors: "It may seem strange to put the author's own name at the top of the list to whom we wish to provide the reproducible research, but it often seems that the greatest beneficiary of preparing the work in a reproducible form is the original author!" It is equally notable that the public was listed last; all of the other constituencies mentioned would likely exist within the small orbit of an individual investigator~\citep{claerbout2001cd}. In Claerbout's discussion, the primary focus is on improving the transparency and productivity of the lab itself, given that much time can be lost attempting to re-create past findings for the sole purpose of understanding what was previously done.

Buckheit and Donoho introduced much of the statistical community to the concept of reproducibility with an influential paper in 1995 detailing their WaveLab software for implementing wavelets for data analysis~\citep{buck:dono:1995}. Citing Claerbout as a strong influence, their rationale for promoting reproducible research produced a useful summary of Claerbout's ideas that has since be repeated many times:
\begin{quote}
An article about computational science in a scientific publication is not the scholarship itself, it is merely advertising of the scholarship. The actual scholarship is the complete software development environment and the complete set of instructions which generated the figures.
\end{quote}
The general conclusion was that delivering a research end product such as a figure or table was no longer sufficient. Rather, the software environment and the means to create the end product must also be delivered, as those additional elements represent the actual scholarship. In order to satisfy this requirement, one would have to make available the \textbf{data} and the \textbf{computer code} used to generate the results.

\section{Reproducible Research}

The definition of reproducible research generally consists of the following elements. A published data analysis is reproducible if the analytic datasets and the computer code used to create the data analysis is made  available to others for independent study and analysis~\citep{peng:2011}. This definition is sufficiently vague that it ultimately raises more questions than it answers. What is an ``analytic dataset"? What does it mean to be ``available"? What is included with the ``computer code"?

Published research can be thought of as living on a continuum up until the point of publication~\citep{peng:domi:zege:2006}. Starting from question formulation and study design, proceeding to to data collection, data processing,  data analysis, and finally to presentation. Along this journey, various elements are introduced to aid in executing the research, such as computing environments, measurement instruments, and software tools. One could choose to make available to others any aspect of this sequence, depending on the practicalities of doing so and the relevance to the final published results. It is challenging to develop a universal cut point for determining which aspects of an investigation should be disseminated and which are not required. However, within various research communities, internal standards have developed and are continuously evolved to keep pace with technology~\citep[e.g.][]{brazma2001minimum,sandve2013ten}.

The analytic dataset generally contains all of the data that can be directly linked to a published result or number. For example, if a paper publishes an estimate of the rate of hospitalization for heart attacks, but the overall study also collected data on hospitalizations for influenza, the influenza data may not be part of the analytic dataset if it makes no appearance in the published result and is not otherwise relevant. While outside investigators may interested in seeing the influenza data (and the original authors may be happy to share it), it is not needed for the sake of reproducible research. The analytic computer code is any code that was used to transform the analytic dataset into results. This may include some data processing (such as variable transformations) as well as modeling or visualization. Generally, the software environment in which the analysis was conducted (e.g. R, Python, Matlab) does not need to be distributed if it is easily obtainable or open source. However, niche software which may be unfamiliar to many readers may need to be bundled with the data and code. 


Upon first consideration, many see reproducibility as a non-issue. How could it be that applying the original computer code to the original dataset would \textit{not} produce the original results? The practical reality of modern research though is that many even simple results depend on a precise record of procedures and processing and the order in which they occur. Futhermore, many statistical algorithms have many tuning parameters that can dramatically change the results if they are not inputted exactly the same way as was done by the original investigators~\citep{haibe2020importance}. If any of these myriad details are not properly recorded in the code, then there is a significant possibility that the results produced by others will differ from the original investigators' results.

\subsection{A Sidebar on Terminology}

The terminology of reproducible research can be bewildering to some in the scientific community because there is little agreement about the meaning of the phrase in relation to other related concepts~\citep{barba2018terminologies,goodman2016does,plesser2018reproducibility}. In particular, one related concept with which all scientists are concerned is what we refer to here as \textit{replication}. In this review, we define replication as the independent investigation of a scientific hypothesis with newly collected data, new experimental setups, new investigators, and possibly new analytic approaches. In a thorough investigation of the terminologies of reproducible research, Lorena Barba found that some fields of study made no distinction between ``reproducible'' and ``replicable'' while some fields used those terms to mean the exact opposite of how we define them here~\citep{barba2018terminologies,national2019reproducibility}. However, a significant plurality of fields, including epidemiology, medicine, and statistics appear to adopt the definitions we use here. 

A key distinction between reproducibility and replication is that reproducibility does not allow for any real variation in the results. If an independent investigator were to \textit{reproduce} the results of another investigator with the original data and code, there should not be any variation between the two investigators' results, except for some allowance for differences in machine precision. Thus, exact reproducibility is sometimes referred to as ``bitwise reproducibility"~\citep{national2019reproducibility}. However, \textit{replication} generally allows for differences in results that arise from statistical variability. Two independent investigators conducting the exact same experiment should, in theory, only differ by an amount quantified by the standard deviation of the data. More generous definitions of replication allow for slightly different study designs, analytic populations, or statistical techniques~\citep{national2019reproducibility}. In those cases, differences in results may arise beyond simple statistical variation. Patil~et al.~have devised a useful visualization of what may or may not differ when reproducing or replicating a published study~\citep{patil2019visual}. The relationship between reproducible research and replication is a topic to which we will return in greater detail in Section~\ref{sec:goals}.

It is difficult to argue that interest in exactly reproducing another investigator's work is anything but a modern phenomenon~\citep{drummond2018reproducible}. Interest in reproducibility prior to the computer and internet age was likely low or non-existent given that there was generally no expectation that investigators would share data in papers---there was simply no practical way to do that except for very small datasets. In the past, other investigators could only resort to independently replicating a published study using their own data collection and whatever high-level description of the methods that was available in the paper. In this setting, detailed descriptions of the methods of analysis were critical if others were to execute the same approach. If the process of conducting the experiment or analysis was simple enough or were sufficiently standardized, then it could be reasonably described in the confines of a journal paper. Suggesting that analyses be described with data and code is a departure from previous ways of communicating scientific results, which relied on describing experiments and analyses in more general terms to give readers the highlights of what was done. A more abstract approach could not be taken with this new form of computational research because the proper abstractions for communicating ideas and standardization of approaches were not yet available.

\subsection{Newer Developments}

The concept of reproducible research was developed to achieve arguably modest goals. Its original aims focused on providing an approach to better communicate the details of computationally intensive research to one's collaborators, colleagues, students and oneself. But two key developments over the past 30 years have changed the context around which reproducible research lives. Although the definition of reproducibility has not changed much since the 1990s, almost everything else about scientific research has. 

In much of the early literature on reproducible research, the focus is on ``computationally-intensive" research which, because of its reliance on complex computer algorithms, was considered perhaps more impenetrable than other research. Fast forward 30 years and the use of computing in scientific research is ubiquitous. It is no longer the domain of niche geophysical scientists or mathematical statisticians using  obscure computer packages. Now, all scientific research involves the use of powerful computers, whether it is for the data collection, the data analysis, or both. Furthermore, the increase in complexity of statistical techniques over this time period has resulted in the need for detailed descriptions of analytic approaches and data processing pipelines. We are all computational scientists now and as a result the concept of reproducibility is relevant to all scientists.

Along with computing power, another key advance over the past 30 years has been the development of the internet. Claerbout's original scheme for distributing data and code to others was via CD-ROM disks, which was a perfectly reasonable approach at the time. However, the need for a physical medium greatly limited the transferrability of information to a large audience. With the development of the internet, it became possible for academics to distribute data and code to the entire world for seemingly minimal cost. This increase in distribution reach changed the nature and importance of reproducible research from primarily improving the internal efficiency of the lab to allowing others to anonymously build on one's own work. The internet dramatically grew the size of an investigator's personal scientific community to include many members beyond one's immediate circle of collaborators. This phenomenon has provided significant benefits to science but there are some implications that could threaten the viability of maintaining and supporting reproducible research in the long-run. Before we consider these implications, we must first consider what are the goals of reproducible research and what problems we want reproducibility to solve.

\section{The Goals of Reproducible Research}
\label{sec:goals}

Beyond communicating the details of an investigation, what are the goals of making research reproducible? The stated goals achieved by making research reproducible have evolved over time since the early 1990s and have become somewhat more elusive. Originally, the goal was to better reveal the process of doing the research. Computational research added a new complexity in the form of software code and high-dimensional datasets and that complexity made understanding the research process more difficult for a reader to infer. Therefore, the solution was to simply publish every step in the process along with the data. Claerbout and colleagues were concerned that others (including themselves!) would not be able to \textit{learn} from what they have done if they did not have the details. The easiest way to do this was via the literal computer code that executed the steps. Any less precise format could risk omitting a key step that affected downstream results~\citep{haibe2020importance}.

Reproducible research comes with a few side benefits. In addition to being able to fully understand the process by which the results were obtained, readers also get the data and the computer code, both of which are valuable to the extent that they can be re-used or re-purposed for future studies or research. Some have suggested that making data and computer code available to others is a \textit{per se} goal of reproducible research, because both can be built upon and leveraged to further scientific knowledge~\citep{gent:temp:2007}. However, such an interpretation is an extension of the original ideas of reproducibility. The former view saw data and code as a medium for communicating ideas whereas the latter view sees data and code essentially as \textit{products} or \textit{digital research objects} to be used by others~\citep{stodden2015reproducing}. While converting a dataset into a data product and packaging computer code into usable software may seem like nominal tasks given that the underlying data and code already exist, there are non-negligible costs associated with the development, maintenance, and support of these products.

Another goal of reproducible research is to provide a kind of audit trail, should one be needed. In fact, one could suggest a definition of reproducible research as ``research that can be checked". Desiring an audit trail for data analyses raises the question of when such an audit trail might be used? In general, one might be interested in seeing the details of the data and the code for an analysis when there is a curiosity about how a specific result was reached. Sometimes that curiosity is raised because of suspicion of an error in the analysis, but other times there is a desire to learn the details of new techniques or methodologies~\citep{haibe2020importance}. Thus, reproducibility primarily concerns the integrity and transparency of the \textit{data analysis} for an investigation. Unlike replication, reproducibility allows for an internal check on the results and is not immediately connected to the context of the outside world.

\section{Reproducibility and Data Analysis}

One could summarize the goal of reproducible research as providing a means to answer the question, ``Do I understand and trust this data analysis?" With the computational nature of today's research, we cannot hope to answer that question without being able to look at the data and the code. In addition, we may wish to know things about the experimental or study design as well as the hypothesis being examined~\citep{hick:peng:2019}. Given the claimed results, the data, and code, one can theoretically determine the reproducibility status of a data analysis.
Reproducibility gives us the means by which we can assess our confidence and trust in an analysis, but it is important to reiterate that the mere fact of reproducibility of an analysis is not a check on validity of the analysis.

The notion of reproducibility as a binary or perhaps multi-level ``state" of a data analysis is a useful characterization in part because it is one of a few qualities of a data analysis that can be immediately verified. Unlike with replication, we do not need to wait for future studies to be conducted in order to determine the reproducibility of an analysis. However, this suggests reproducibility's usefulness is limited. What do we ultimately learn from merely reproducing the results of an analysis? For example, it may be possible to execute code on a dataset without ever looking at the code or the data. In that case, the original goal of reproducibility---to learn about the details of an investigation---has been thwarted. We have simply learned that the code produces what the authors claim the code produces. In general, executing a process and seeing that process produce the results exactly as they were expected, produces very little new information. 

The statement of the question ``Do I understand and trust this data analysis?" depends critically on the perspective of the person asking the question. If the person asking is an expert in the area, they might be able to glance at the code and data and understand immediately what is going on. A non-expert in the field might be able to execute the code and produce results without ever understanding the operations of the analysis. An adjacent question that might be worth asking is ``Is this data analysis understandable and trustworthy?" However, this question is not any easier to answer because it hypothesizes underyling objective qualities of a data analysis. But opinions may still vary widely about what these underlying qualities should be depending on who is asking the question. To answer either question, one needs to look carefully at the data and the code to learn exactly what was done. But ultimately, the data and code only represent a piece of the answer. Whether an analysis can be understood or trusted depends critically on many aspects outside of the analysis itself, including the perspective of the person reading the analysis.

Nevertheless, one hope is that reproducibility can lead to higher quality data analyses. The logic is that requiring all analyses to provide data and code would put investigators on notice that their work would be scrutinized. However, one high-profile example suggests this is unlikely to be the case.

\subsection{Example: Forensic Bioinformatics}

In a now-retracted 2006 study by Potti~et~al.~\citep{potti2006genomic}, the investigators claimed to have identified genomic signatures using microarrays that could predict whether an individual responded well to chemotherapy. The analysis was conducted using data from publicly available cell lines, and so the data were in a sense available. However, subsequent attempts to reproduce the findings failed and reproducibility was only achieved when errors were deliberately inserted into the analysis code~\citep{coom:wang:bagg:2007,baggerly2009deriving}. Keith Baggerly, Kevin Coombes, and Jing Wang meticulously reconstructed the error-prone analysis and laid out all of the details in both text and code. Ironically, they were ultimately able to reproduce the analysis of Potti~et~al. after significant reverse engineering and forensic investigation. In fact, we might never have learned what mistakes were made if Baggerly and his colleagues were not able to reproduce the analysis.

The example of the Potti study is a pathological example of a reproducible analysis (after much forensic investigation) being profoundly incorrect. However, it is worth asking what role reproducibility \textit{might have played} in this case? If Potti et al.~had released the code and data that were clearly linked together, perhaps as a research compendium~\citep{gent:temp:2007}, then the errors could have been found more quickly. However, given the sheer number and complexity of the problems, it still likely would have taken some time to understand them all. Coombes~et~al. published their letter only a year after the initial publication, so the timeline might have been advanced by a few months. However, a key fact would remain---the flawed analysis was already completed. Furthermore, once the truth was ultimately revealed to the authors, it took years of further investigation by many others before the original paper was retracted.

\subsection{Reproducibility and Quality: A Prevention Model?}

Examples like the Potti paper raise the question of whether demanding or requiring reproducibility of a study beforehand can pre-emptively improve its quality. Evidence of this connection between reproducibility and quality is lacking, which is not surprising given that the question is somewhat ill-posed. What exactly are we looking for in a ``high-quality" data anlaysis? One could hypothesize that if an investigator knew in advance that the data and the code would be publicly available for scrutiny, then they would take the extra effort to make sure that the analyses were properly done. Perhaps if Potti~et~al. had been forced to make their code publicly available, they would have checked it first.

In the case of Potti~et~al.~we now know that requiring reproducibility or even just code sharing would not have made much difference. Reporting done by \textit{The Cancer Letter} showed definitively that the investigators were aware of numerous statistical and coding errors with the analysis but did not think they were serious problems~\cite{goldberg:2015}. Rather, they were considered ``differences of opinion". The notion that requiring reproducibility can lead to improved data analyses relies on the critical assumption that the investigators are able to recognize what is an error in the first place. If they do recognize the error and hide it, then that is fraud. If they do not recognize the error and publish it anyway, then that is at best careless. However, in both cases, forcing the data and code to be published would not have made any difference.

\subsection{Replication and Reproducibility}

Reproducibility does not provide a useful route to preventing poor data analyses from occurring, but it does provide the basis for a meaningful discussion about whether their might be problems in the analysis and how such problems might be fixed. Replication differs from reproducibility primarily because it addresses a different goal. Replication answers the question, ``Is this scientific claim true?" Reproducibility addresses the integrity of the data analysis that generated the evidence for a scientific claim, while replication addresses the integrity of the claim itself in the context of the outside world. Fundamentally, reproducible research has relatively little to say on the question of external validity. Claims resulting from reproducible results can be both correct and incorrect~\citep{leek:peng:2015}. Claims resulting from irreproducible results are less likely to be true, but that may depend on the reasons for the lack of reproducibility. For example, evidence generated via random algorithms may not be exactly reproducible if random number generator seeds are not saved, but the underlying evidence may still be sound. Ultimately, claims made by irreproducible studies may in fact be true, but irreproducible studies simply do not provide evidence for such claims.

\subsubsection{Example: Re-analysis of Air Pollution Studies}

In the mid 1990s, two large studies of ambient air pollution and mortality---The Six Cities Study~\citep{dockery1993association} and the American Cancer Society (ACS) Study~\citep{pope1995particulate}---were published, presenting evidence that differences in air pollution concentrations between cities were significantly associated with rates of mortaliy in these cities. Both studies came under intense scrutiny when the U.S. Environmental Protection Agency cited the results in their revision of the National Ambient Air Quality Standards for fine particles. In particular, there were demands from numerous corners that the data used in the studies should be made available. However, the data in these studies, as with most health-related studies, included personal information about the subjects and arguments were made that promises of confidentiality had to be kept. To address the impasse of making the data available, the original investigators engaged the Health Effects Institute (HEI) to serve as a kind of trusted third party to broker a reanalysis of the studies. Ultimately, HEI recruited a research team lead by investigators at the University of Ottawa to obtain the original data for both the Six Cities and ACS studies, reproduce the original findings, and conduct additional sensitivity analyses to assess the robustness of the original findings~\citep{krewski2000reanalysis}.

The extensive reanalysis found that the original studies were largely reproducible, if not perfectly reproducible. For the Six Cities study, the key result was a mortality relative risk of $1.26$, which the re-analysis team computed to be $1.28$. For the ACS study the original mortality relative risk was $1.17$, close to the re-analysis value of $1.18$.
While one could argue that these studes were strictly speaking not reproducible, such small differences are not likely to be material. In fact, we now know, after numerous follow-up studies and independent replications, that the core findings of both studies appear to be true~\citep{broo:raja:pope:2010} and that the U.S. EPA itself rates the evidence of a connection between fine particles and mortality to be ``likely causal"~\citep{isa:2009}. The re-analysis team ran many other analyses, including variables that had not been considered in the original studies. Overall, they found that the sensitivity analyses did not change any of the major conclusions. Interestingly, one of the key conclusions of the final report from HEI was that at the end of the day ``No single epidemiologic study can be the basis for determining a causal relation between air pollution and mortality"~\citep{krewski2000reanalysis}.

The HEI re-analysis of the Six Cities and ACS studies highlights the role of trust in data analysis. Prior to the re-analysis, many parties simply did not trust that the analysis was done properly or that all reasonable competing hypotheses had been considered. While making the data available might have allowed others to build that trust for themselves, allowing a neutral third party to examine the data and reproduce the findings at least ensured that one other group had seen the data. In addition, HEI's role in organizing the expert panel, conducting public outreach, and managing an open process played an important role in building trust in the community. While not all parties were completely satisified with the process, what the re-analysis did was allow fellow scientists to learn from the original studies and gain insight into the process that lead to the original findings. Ultimately, the key goals of reproducible research were achieved.

In hindsight, another lesson learned from the HEI re-analysis is that the importance of reproducibility of a given study can fade with time. Over 25 years later, there have been scores of follow-up studies and replications that have largely come to similar conclusions as the Six Cities and ACS studies. Although both studies remain seminal in the field of air pollution epidemiology, they could be deleted from the literature at this point and have little impact on our understanding of the science. This is not to say that the data and ongoing analyses do not have value, but rather the original results have been subsumed by later studies. Reproducibility was only critical when the studies were first published because of the paucity of large studies at the time.

\section{Reproducibility and Better Data Analysis}

Recent work has focused on the quality and variability of data analyses published in various fields of study~\citep[e.g.][]{open2015estimating,jager2014estimate,ioannidis2005most,patil2016should}, with some claiming the existence of a ``replication crisis" due to the wide variation between studies examining the same hypotheses~\cite{schooler2014metascience}. The causes of this variation between studies are myriad but one large category includes various aspects of the data analysis. Because of the increasing complexity of data analyses, many choices and decisions must be made by analysts in the process of obtaining a result. With these increasing complexities we also increase the risk of human error and bias in data analysis. These choices and decisions often have an unknown impact on the final estimates produced and therefore may or may not be recorded by the investigators~\citep{stodden2015reproducing}. These ``research degrees of freedom'' allow investigators to unknowingly, or perhaps knowingly, steer data analyses in directions that may support specific hypotheses rather than represent all of the evidence in the data~\citep{simmons2011false}.

What role can reproducible research play in improving the quality of data analyses across all fields? The answer can be found in part with the experience of the HEI re-analysis of the Six Cities and ACS air pollution studies. Because they were re-analyses, one could imagine the expectation was that the results would be confirmed to some reasonable degree. If there was a significant deviation from the published results, then we would have to dig into the original analysis to discover why. Because the results were largely reproduced, one could argue that little was learned. However, additional analyses were done and sensitivity analyses were conducted. As a result, we learned much about the data analysis process. The re-analysis thus produced valuable knowledge about how to analyze air pollution and health data. 

For example, the re-analysis team noted that both mortality and air pollution were highly spatially correlated, a feature that was not considered in the original analysis. They noted, ``If not identified and modeled correctly, spatial correlation could cause substantial errors in both the regression coefficients and their standard errors. The Reanalysis Team identified several methods for dealing with this, all of which resulted in some reduction in the estimated regression coefficients."~\cite{krewski2000reanalysis} In addition, reproducibility helps free up time for the analysts interested in re-analyzing the data to focus on parts of the data analysis that require more human intepretation. For example, if an independent data analyst knew that an analysis was already reproducible, then more time and resources would be available to understand {\it why} a specific model was chosen, instead of {\it what} version of software was used to run this model. In the re-analysis of the data from Potti et al.~\citep{potti2006genomic}, Baggerly and Coombes noted that they had spent thousands of hours re-examining the data attempting to reproduce the original results~\citep{baggerly2010disclose,goldberg2014duke}.

There are also different degrees of reproducibility when building a data analysis and differences in audiences that may or may not be allowed to have access to these components. For example, a data analyst may choose to make the data available, but not the code (or the opposite). Others may make both the code and data available for only one audience (Audience A), but not for another audience (Audience B). There are valid reasons why an analyst might choose to do this, such as if the data analysis uses data with protected health information in a hospital setting or if the data analyst works at a business or company and cannot share the code or data with others outside of the company. It is important to note that just because an analysis is not fully reproducible to one audience (Audience B) does not mean that it is an invalid analysis with incorrect conclusions. While it does make it harder for Audience B to trust the results, it still can be a valid or correct analysis. However, the lack of reproduciblity to this audience may mean that the evidence supporting any claims is weaker. Despite these potential differences in degrees of reproducibility, as demonstrated in the HEI re-analysis, efforts made to make a data analysis more reproducible is a step in the right direction for making it a better data analysis.

Ultimately, the reproducibility of research, when possible, allows us a significant opportunity to (1)~learn from others about how best to analyze certain types of data; (2)~reduce human errors and bias as data become larger and more complex; (3)~free up time for re-analyzers to focus on parts of a data analysis that require more human interpretation; (4)~have discussions about what makes for a good data analysis in certain areas of study; and (5)~improve the quality of future data analyses. When teaching data analysis to students, it is common to talk in abstractions and theories, describing statistical methods and models in isolation.  When real data is shown, it is often in the form of toy examples or in short excerpts. Increasing the reproducibility of all studies presents an opportunity to dramatically expand instruction on the craft of data analysis so that core set of elements and principles for characterizing high quality analyses can be established within a field~\citep{hick:peng:2019}.

\section{Refining Reproducibility}

In the thirty years since the idea of reproducible computational research was brought the forefront of the research community, we have learned much about its role and its value in the research enterprise. The original goal of providing a transparent means by which researchers can communicate what they have done and allow others to learn remains a primary rationale. Reproduciblity has a secondary role to play in improving the quality of data analysis in that it serves as the foundation on which people can learn how others analyze data. Without code and data, it is nearly impossible to fully understand how a given data analysis was actually done. But much about computational research has changed in the past 30 years and we can perhaps develop a more refined notion about what it means to make research ``reproducible". The two key ideas about reproducibility---data and code---are worth revisiting in greater detail.

\subsection{Data}

The sharing of data is ultimately valuable in and of itself. Data sharing, to the extent possible, reduces the need for others to collect similar data, allows for combined analyses with other datasets, and can create important resources for unforeseen future studies. Datasets can retain their value for considerable time, depending on the area and field of study. One example of the value of data sharing comes from the National Mortality, Morbidity, and Air Pollution Study, a major air pollution epidemiology study conducted in the late 1990s and early 2000s~\citep{nmmaps1,nmmaps2}. The mortality data for this study were shared on a web site and then later updated with new data. A systematic review found 67 publications had made use of the dataset, often to demonstrate the development of new statistical methodology~\citep{barnett2012benefits}. In addition, the release of the data at the time allowed for a level of transparency and trust in air pollution research that was novel for its time. 

Today, many data sharing web repositories exist that allow easy distribution of data of almost any size. While in the past, an investigator interested in sharing data had to purchase and setup a web server, now investigators can simply upload to any number of services. The Open Science Framework~\citep{foster2017open}, Dataverse Project~\citep{king2007introduction}, ICPSR~\citep{swanberg2017inter}, and SRA~\citep{leinonen2010sequence} are only a handful of public and private repositories that offer free hosting of datasets. The major benefit of repositories such as these is to absorb and consolidate the cost of hosting data for possibly long periods of time.

The view of data sharing as inherently valuable is not without its challenges. Indeed, stripping data from its original context can be problematic and lead to inappropriate ``off-label" re-use by others. It has been argued that data only has value in its explicit connection to the knowledge that it produces and that we must be careful to preserve the connections between the data and the knowledge they generate~\citep{stodden2020beyond}.

Recently, best practices for sharing data have been developed. Some of these practices are specific to areas of study while some are more generic. In particular, the emergence of the concept of \textit{tidy data} has provided a generic format for many different types of data that serves as the backbone of a wide variety of analytic techniques~\citep{wickham2014tidy}. Practical guidance on sharing data via commonly used spreadsheet formats~\citep{broman2018data} and on providing relevant metadata to collaborators is now widely applicable to many kinds of data~\citep{ellis2018share}.

\subsection{Code}

The primary role of sharing code is to communicate what was done in transforming the data into scientific results. Today, almost all actions releveant to teh science will have occurred on the computer and it is essential that we have a precise way to document those actions. Computer code, via any number of programming and data analytic languages, is the most precise way to do that. The sharing of code generally represents less of a technical burden than the sharing of data. Code tends to be much smaller in size than most datasets and can easily be served by code sharing services such as GitHub, BitBucket, SourceForge, or GitLab. 

While the benefits of code sharing tend to focus on the code's usability and potential for re-purposing in other applications, it is important to reiterate that code's primary purpose is to communicate what was done. In short, code is not equivalent to \textit{software}. Software is code that is specifically designed and organized for use by others in a wide variety of scenarios, often abstracting away operational details from the user. The usability of software depends critically on aspects like design, efficiency, modularity, and portability---factors that should not generally play a role when releasing research code. Sharing research code that is poorly designed and inefficient is far preferable to not sharing code at all. That said, this  notion does not preclude the possibility for best practices in developing and sharing research code.

Software is often a product of research activity, particularly when new methodology is developed. In those cases, it is important that the software is carefully considered and designed well for its intended users. However, it should not be considered a requirement of reproducible research that software be a product of research. For software that is developed for distribution, there is increasing guidance for how such software should be distributed. Software package development has become easier for programming languages like R, which have robust developer and user communities~\citep{r2020}, and numerous tools have been developed to make incorporating code into packages more straightforward for non-professional programmers~\citep{devtools2020}. In addition, the concept of testing and test-based developed has been shown to be a useful framework for setting expectations for how software should perform and identifying errors or bugs~\citep{testthat2011}.

\section{Future Directions}

Technological trends over time generally favor a more open approach to science as the costs of sharing, hosting, and publishing have gone down. The continuing rapid advancement of computing technology, internet infrastructure, and algorithmic complexity will likely introduce new challenges to reproducible research. As the scientific community expands its sharing of data and code there are some important issues to consider going forward.

The rapidly evolving nature of scientific communication serves to highlight the role of reproducibility in advancing science. Without reproducibility, countless hours could be wasted simply trying to figure out what was done in a study. In situations were key decisions must be made based on scientific results, it is important that the robustness of the findings can be assessed quickly without the need for guessing or inferences about the underlying data. A stark example can be drawn from the COVID-19 pandemic. In April 2020 little was known about the disease and a study was published on medR$\chi$iv producing an estimate of the prevalence of COVID-19 in the population~\citep{bendavid2020covid}. At the time, important public health decisions had to be made in response to the pandemic and any information about the disease would have been highly relevant. Upon publication, numerous criticisms about the study's design and analysis appeared on social media and the web. However, the aspect most relevant to this review is that in many of the critiques, substantial time was taken to simply guess at what the researchers had done. Although a written statistical appendix was provided with the paper, no data or code were published along with the study. As a result, independent investigators had little choice but to infer what was done.

The urgency of decision-making based on scientific evidence can exist in a variety of situations, not just on the the minute-by-minute timescale of a worldwide pandemic. Many regulatory decisions in environmental health have to be made based on only a handful of studies. Often, there is no time to wait years for another large cohort study to replicate (or not) existing findings. In such situations where decisions need to be made, the more code and data that can be made available to assess the evidence, the better. In the interim, followup studies can be conducted and revisions to the evidence base can be made in the future if needed. The re-analyses Six Cities and ACS studies provide a clear example of this process and history has shown those results to be highly consistent across a range of replication studies.

The maintenance of code and data is generally not a topic that is discussed in the context of reproducible research. When a paper is published, it is sent to the journal and is considered ``finished'' by the investigators. Unless errors are found in the paper, one generally need not revisit a paper after publication. However, both code and data need to be maintained to some degree in order to be useful. Data formats can change and older formats can fall out of  favor, often making older datasets unreadable. Code that was once highly readable can become unreadable as newer languages come to the fore and practitioners of older languages decrease in number. Maintenance of data and code is not a question of paying for computer hardware or services. Rather, it is about paying for people to periodically update and fix problems that may be introduced by the constantly changing computing environment.

Unfortunately, funding models for scientific research are aligned with the mechanism of paper publication, where one can definitively mark the end of a project (and also the end of the funding). However, with data and code, there is often no specific end point because other investigators may re-use the data or code for years into the future. Term-based project funding, which is the structure of almost all research funding, is simply not designed to provide support for maintaining materials on an uncertain timeline.

The first thirty years of reproducible research largely centered on discussions of the validity of the idea and what value it provided to the scientific community. Such discussions are largely settled now and both data and code share are practiced widely in many fields of study. However, we must now engage in a second phase of reproducible research which focuses on the continued development of infrastructure for supporting reproducibility.

\clearpage

\bibliography{combined}
\bibliographystyle{asa}

\end{document}